\documentclass[fleqn,10pt,table]{wlscirep}
\usepackage[utf8]{inputenc}
\usepackage[T1]{fontenc}
\usepackage{bm}
\usepackage{float}
\usepackage{multicol}
\usepackage{multirow}
\usepackage{marvosym}
\usepackage{hyperref}
\definecolor{myblue}{cmyk}{77,55,0,0}
\definecolor{myorange}{cmyk}{3,63,87,0}
\usepackage{wrapfig}

\definecolor{airforceblue}{rgb}{0.36, 0.54, 0.66}

\title{PhyloTransformer: A Discriminative Model for Mutation Prediction Based on a Multi-head Self-attention Mechanism}
\author[1,\dag,\Letter]{Yingying Wu}
\author[2,\dag]{Shusheng Xu}
\author[1,3,\Letter]{Shing-Tung Yau}
\author[2, \Letter]{Yi Wu}
\affil[1]{Harvard University, Center of Mathematical Sciences and Applications, Cambridge, U.S.}
\affil[2]{Tsinghua University, Institute for Interdisciplinary Information Sciences, Beijing, China}
\affil[3]{Harvard University, Department of Mathematics, Cambridge, U.S.}
\affil[$\dag$]{These authors contributed equally}
\affil[\Letter]{e-mail: ywu@cmsa.fas.harvard.edu}
\affil[\Letter]{e-mail: yau@math.harvard.edu}
\affil[\Letter]{e-mail: jxwuyi@mail.tsinghua.edu.cn}

\begin{abstract}

Coronaviruses are enveloped non-segmented positive-sense RNA viruses. Severe acute respiratory syndrome coronavirus 2 (SARS-CoV-2) has caused an ongoing pandemic infecting 219 million people as of October 19, 2021, with a 3.6\% mortality rate. Although coronaviruses have RNA proofreading functions, a large number of variants still exist as quasispecies. Natural selection can generate favorable mutations with improved fitness advantages, including pathogenicity, infectivity, transmissibility, angiotensin-converting enzyme 2 (ACE2) binding affinity, and antigenicity. However, the identified coronaviruses might just be the tip of the iceberg, and potentially more fatal variants of concern (VOCs) may emerge over time. Understanding the patterns of emerging VOCs and forecasting mutations that may potentially lead to gain of function or immune escape is urgently required. Here we developed PhyloTransformer, which is a Transformer-based discriminative model that engages a multi-head self-attention mechanism to model genetic mutations that may lead to viral reproductive advantage. In order to identify complex dependencies between the elements of each input sequence, PhyloTransformer utilizes advanced modeling techniques, including a novel Fast Attention Via positive Orthogonal Random features approach (FAVOR+) from Performer, and the Masked Language Model (MLM) from Bidirectional Encoder Representations from Transformers (BERT). PhyloTransformer was trained with 1,765,297 genetic sequences retrieved from the Global Initiative for Sharing All Influenza Data (GISAID) database. Firstly, we compared the prediction accuracy of \textit{novel mutations} and \textit{novel combinations} using extensive baseline models, including a Transformer-based local model, called Local Transformer, and other local models, such as ResNet-18, multilayer perceptron, logistic regression, KNN, random forest, and gradient boosting; we found that PhyloTransformer outperformed every baseline method with statistical significance. Secondly, we examined predictions of mutations in each nucleotide of the receptor binding motif (RBM), which is a specific sequence of amino acids from the SARS-CoV-2 spike protein that mediates the binding of spike protein to ACE2. Our predictions displayed preciseness and accuracy: our model predicted a total of two mutations in the RBM, and these two mutations precisely coincided with two of the four important mutations presented in seminal bench studies. Thirdly, we predicted modifications of N-glycosylation sites to help identify mutations associated with altered glycosylation that might be favored during viral evolution. We anticipate that the viral mutations predicted by PhyloTransformer may shed light on potential new mutations that may lead to fitness advantages of SARS-CoV-2 variants. Thus, our predicted variants may guide therapeutics and vaccine design for effective targeting of future SARS-CoV-2 variants.
\end{abstract}

\begin{document}
\flushbottom
\maketitle
\thispagestyle{empty}

\section*{Introduction}

Severe acute respiratory syndrome coronavirus 2 (SARS-CoV-2) is the causative agent of Coronavirus disease 2019 (COVID-19). The unprecedented COVID-19 pandemic is one of three major pathogenic zoonotic disease outbreaks caused by $\beta$-coronaviruses in the past two decades \cite{cui2019origin,de2016sars}. Severe acute respiratory syndrome coronavirus (SARS-CoV) emerged in 2002, infecting 8,000 people with a 10\% mortality rate \cite{patel2020orthopaedic, hui2020continuing}. Middle East respiratory syndrome coronavirus (MERS-CoV) emerged in 2012 with 2,300 cases and a 35\% mortality rate \cite{graham2010recombination}. The third outbreak, mediated by SARS-CoV-2, emerged in 2019 with a mortality rate of 3.6\% \cite{baud2020real} and 219 million cases have been reported as of October 2021.

After the emergence of SARS-CoV-2 in late 2019, the virus exhibited relative evolutionary stasis for approximately 11 months. Since the end of 2020, SARS-CoV-2 has consistently acquired approximately two mutations per month \cite{worobey2020emergence, duchene2020temporal} resulting in novel variants of concern (VOCs).
As more individuals became vaccinated against SARS-CoV-2, the viral evolution has been characterized by the emergence of sets of mutations, probably in response to the changing immune profile of the human population. Currently, the main focus is to identify critical SARS-CoV-2 countermeasures, including vaccines, therapeutics, and diagnostics.

Since coronaviruses have proofreading functions \cite{smith2013coronaviruses}, most mutations in the SARS-CoV-2 genome are expected to comprise neutral amino acid changes with little or no impact on fitness advantages \cite{maclean2020no}. However, the evolutionary diversity introduced by a small minority of mutations may impact the virus phenotype and promote virus fitness. Some of the SARS-CoV-2 mutations displayed positive selection with improved pathogenicity, infectivity \cite{ yurkovetskiy2020structural}, transmissibility \cite{hou2020sars, volz2021evaluating}, angiotensin converting enzyme 2 (ACE2) binding affinity \cite{starr2020deep}, or antigenicity \cite{thomson2021circulating}. In addition, other SARS-CoV-2 mutations introduced an optimized trade-off to improve overall fecundity. Heavily mutated lineages have also been reported, such as the lineage B.1.1.298, which harbors the following four amino acid substitutions: $\Delta$H69–V70, Y453F, I692V, and M1229I \cite{fonager}. Some mutations may amplify other mutations, providing an improved fitness advantage. For example, the combination of E484K, K417N, and N501Y results in the highest degree of conformational alterations compared to either E484K or N501Y alone \cite{nelson2021molecular}. Accumulating evidence suggests that mutations which require immediate attention are circulating, which highlights the urgent need to develop effective prevention and treatment strategies.

While vaccination has been the most important and effective preventive measure, it is also facing challenges. The mRNA vaccine BNT162b2 (Pfizer–BioNTech) has 95\% efficacy against COVID-19 \cite{polack2020safety}. However, the estimated effectiveness of the vaccine against the B.1.1.7 variant was 89.5\% (95\% CI, 85.9 to 92.3) at 14 or more days after the second dose and 75.0\% (95\% CI, 70.5 to 78.9) against the B.1.351 variant \cite{abu2021effectiveness} at 14 or more days after the second dose. Several studies have characterized multiple mutations that change the antigenic phenotype. Thus, these studies elucidate how these mutations affect antibody-mediated neutralization. Variants containing these mutations are potentially highly virulent and have received much recent attention. However, it remains unknown whether more infectious variants exist along with the likelihood that they will appear and transmit. Designing vaccines after a novel variant has emerged is not optimal because the variant could potentially compromise existing vaccines and spread among the population. Thus, more infections might generate further variants, leading to a never-ending pandemic.

In order to win the race against the rapidly evolving SARS-CoV-2, an intelligent system capable of forecasting potential VOCs before they actually appear is urgently required. Therefore, we propose that PhyloTransformer, a novel deep learning model, may be used to predict novel mutations and novel combinations of mutations in SARS-CoV-2. Thus, we anticipate that when variants of high consequence arise, existing vaccines based on PhyloTransformer predictions will have already been developed that target those strains.

\section*{Technical Overview}
Our analysis pipeline was based on the 5/31/21 download of the Global Initiative for Sharing All Influenza Data (GISAID) database (\url{https://www.gisaid.org/}) with a total of 1,765,297 genetic sequences. \rm PhyloTransformer consists of two independently trained models based on the SARS-CoV-2 spike sequence of 1,273 amino acids (3,819 nucleotides).

PhyloTransformer was trained with the full-length spike protein nucleotide sequence based on the Transformer \cite{vaswani2017attention} architecture 
using the Masked Language Model (MLM) pre-training objective in Bidirectional Encoder Representations from Transformers (BERT) \cite{devlin-etal-2019-bert}. The Fast Attention Via positive Orthogonal Random features approach (FAVOR+) from Performer \cite{choromanski2021rethinking} was utilized to accelerate attention computation. Transformer is a sequence model originally proposed for language processing tasks, and the MLM pre-training objective allows for multiple site predictions for phylogenetic applications. Meanwhile, the FAVOR+ technique reduces 
the time complexity for computing intra-genetic interactions that was initially prohibitive with the naive Transformer model.
Employing the FAVOR+ technique permits consideration of global sequential information that accounts for the entire spike sequence. 

PhyloTransformer, which functions as a global model, was trained with the full spike sequence. For comparison of PhyloTransformer to models that only utilize local information, we segmented each spike protein nucleotide sequence into 3,819 sections with a length of 15, while filtering out the repetitions. 
Next, we trained local models using various popular machine learning algorithms, including the standard Transformer architecture and ResNet-18, as well as classical methods including multilayer perceptron (MLP), logistic regression, KNN, random forest, and gradient boosting. We modeled the full spike sequence by integrating FAVOR+ and MLM techniques into PhyloTransformer to reduce both time and spatial complexity to a linear extent; otherwise, the full spike sequence would be computationally prohibitive for Transformer models. PhyloTransformer utilizes the full-length spike sequence as input with multiple masked sites and generates predictions on respective sites simultaneously, while baseline models can only make independent local predictions on a specified site.

We used three datasets based on a temporal cut-off. We first prepared the \it small \rm dataset, which contains sequence data from 01/01/2020 to 03/01/2020 with a total of 24,951 sequences. Next, we prepared the \it medium \rm dataset, which contains sequence data from 01/01/2020 to 11/11/2020 with a total of 134,704 sequences. Finally, we prepared the \it large \rm dataset, which consists of data from 01/01/2020 to 05/31/2021, with a total of  1,765,297 sequences. Each dataset was then evenly split into training and testing sets, retaining their temporal order. Table \ref{tab:voc} summarizes the datasets used in this study.

The term VOC for SARS-CoV-2 is a category used when mutations in the receptor binding domain (RBD) of the viral spike protein substantially increase the binding affinity of the RBD to the human (h)ACE2 receptor, resulting in rapid spread in human populations. Table 1 shows the single amino acid mutations that emerged as VOCs in our \it large \rm dataset. Noteworthily, Table 1 shows an increase in the percentage of VOCs over time. For example, 18.51\% of the sequences in the training set (i.e., the first half of the data) have the mutation N501Y, and in the testing set (i.e., the latter half of the data) the percentage of sequences with the N501Y mutation drastically increased to 73.96\%.
\begin{table}
\begin{center}
\begin{tabular}{|l|r|r|r|r|r|r|r|}
\hline
\hline
\rowcolor{myblue!20}Dataset & Train Start Date & Train End Date & Test End Date & Total & Training Set & Testing Set \\
\hline
\rowcolor{myblue!5}Small &01/01/2020& 03/20/2020 &  03/31/2020 &   24,951 & 12,475 & 12,476\\
\rowcolor{myblue!5}Medium&01/01/2020& 04/22/2020 &  09/30/2020 & 134,704  & 67,352 & 67352 \\
\rowcolor{myblue!5}Large &01/01/2020& 02/17/2021 & 05/31/2021 &  1,765,297 & 882,648 & 882,649\\
\hline
\hline
&\multicolumn{3}{c|}{In Training Set} &\multicolumn{3}{c|}{In Testing Set} \\
\hline\rowcolor{orange!20}VOC
& VOC Mutation & Unmutated & Other Mutation & VOC Mutation & Unmutated & Other Mutation\\
\hline
\rowcolor{orange!10}K417N$^{\beta}$	&   0.81\%  &   99.05\%  &    0.14\%  &    1.67\%  &   95.48\%  &    2.85\%  \\
\rowcolor{orange!10}K417T$^{\gamma}$& 0.14\%  &   99.05\%  &    0.81\%  &    2.85\%  &   95.48\%  &    1.67\%  \\
\rowcolor{orange!10}T478K$^{\delta}$&0.46\%  &   99.48\%  &    0.06\%  &    3.46\%  &   96.48\%  &    0.06\%  \\
\rowcolor{orange!10}L452R$^{\delta}$&  2.38\%  &   97.57\%  &    0.05\%  &    6.41\%  &   93.40\%  &    0.20\%  \\
\rowcolor{orange!10}E484K$^{\alpha,\beta,\gamma}$& 1.60\%  &   98.33\%  &    0.07\%  &    8.78\%  &   90.86\%  &    0.36\%  \\
\rowcolor{orange!10}N501Y$^{\alpha,\beta,\gamma}$&  18.51\%  &   81.24\%  &    0.24\%  &   73.96\%  &   25.89\%  &    0.15\%  \\
\rowcolor{orange!10}D614G$^{\alpha,\beta,\gamma,\delta}$&96.44\%  &    3.54\%  &    0.02\%  &   99.39\%  &    0.60\%  &    0.01\%  \\
\rowcolor{orange!10}P681H$^{\alpha}$&  19.34\%  &   80.30\%  &    0.36\%  &   73.12\%  &   24.39\%  &    2.49\%  \\
\rowcolor{orange!10}P681R$^{\delta}$& 
  0.30\%  &   80.30\%  &   19.41\%  &    2.47\%  &   24.39\%  &   73.14\%\\
\hline
\hline
\end{tabular}
\caption{ \textbf{Datasets used during training.} The analysis was based on the GISAID database. Each dataset was evenly split into training data and testing data while retaining their temporal order. Variant of Concern (VOC) Mutation: the percentage of sequences with VOC mutations. Unmutated: the percentage of sequences that remained the same as the reference sequence at the respective positions. Other Mutation:  the percentage of sequences with single amino acid mutations other than the amino acid mutations that characterize the VOCs.}\label{tab:voc}

\end{center}
\end{table}

\section*{Results}
We used the hCoV-19/Wuhan/WIV04/2019 sequence (WIV04) as our reference sequence, which is the official reference sequence employed by GISAID (EPI\_ISL\_402124). WIV04 represented the consensus of several early submissions for the $\beta$-coronavirus responsible for COVID-19 \cite{okada2020early}, which was isolated by the Wuhan Institute of Virology from a clinical sample of a bronchoalveolar lavage fluid for RNA extraction and metagenomic next-generation sequencing. The consensus sequence was obtained by \it de novo \rm assembly \cite{zhou2020pneumonia}.
Based on WIV04, we define a \it mutation \rm as the change in a nucleotide at a particular position that is different from the reference sequence.
We define a mutation at a particular position that only occurs in the testing set but does not occur within the training set as a \it novel mutation, \rm which signifies a mutation that is novel for the training set. We define all the novel mutations over an RNA sequence as a \it novel combination, \rm i.e., a combination of mutations that do not occur in the training data.  The prediction of \it novel mutations \rm aims to predict single mutations, while the prediction of \it novel combinations \rm aims to predict a collection of single mutations that jointly occur in a mutated sequence.

The prediction accuracies of \it novel mutations \rm and \it novel combinations \rm were evaluated based on 10 checkpoints after the predicting models PhyloTransformer, Local Transformer, and ResNet-18 converged. We first performed lag 1 autocorrelation to test the correlation between accuracy scores obtained from models that are one checkpoint apart. The autocorrelation tests were performed on \it small, medium, \rm and \it large \rm datasets for predicting \it novel mutations \rm and \it novel combinations, \rm with a total of 18 tests.  We found no time dependency between the 10 accuracy scores in each of these 18 tests. For other classical machine learning models, we repeated the experiment 10 times for each dataset. The details are reported in Box 1D.

In this section, we first evaluated PhyloTransformer-generated predictions of \it novel mutations \rm and \it novel combinations. \rm Next, we compared the accuracy of each prediction with those obtained from baseline models. We then reported our predictions in the receptor binding motif (RBM). Finally, we predicted modifications of N-glycosylation sites to help identify mutations associated with altered glycosylation that might be favored during viral evolution. The detailed model architecture and training process are reported in the Methodology section.

\subsection*{Predicting Novel Mutations}
We evaluated the efficacy of PhyloTransformer to predict \it novel mutations \rm and compared it to baseline model predictions from three datasets with different sizes spanning different time frames. The dataset details are described in Table \ref{tab:voc}, and the prediction results are reported in Box 1. For each mutation, we masked the raw nucleotide in the reference sequence and predicted which nucleotide it would mutate to, and we selected the nucleotide with the highest confidence as our prediction. The prediction accuracy is the proportion of positions that are predicted correctly among all novel positions in the testing set. The prediction accuracy of random guessing is exactly 1/3. We evaluated the prediction efficacy averaged over 10 checkpoints after the convergence of PhyloTransformer, Local Transformer, and our baseline models on three datasets with the variance marked either below or above. Next, we reported the model predictions from each dataset, which is displayed in Box 1A.

We performed a two-sample $z$-test of proportions and found that for each model, the best prediction accuracy of \it novel mutations \rm 
from the \it large \rm dataset among the 10 checkpoints significantly less than PhyloTransformer. 
Local Transformer had the best performance among baseline models, but the average over 10 checkpoints was still 11\% lower than that of PhyloTransformer on the \it large \rm dataset with statistical significance, as shown in Box 1D.

\begin{figure}[H]
	\includegraphics[width=\textwidth]{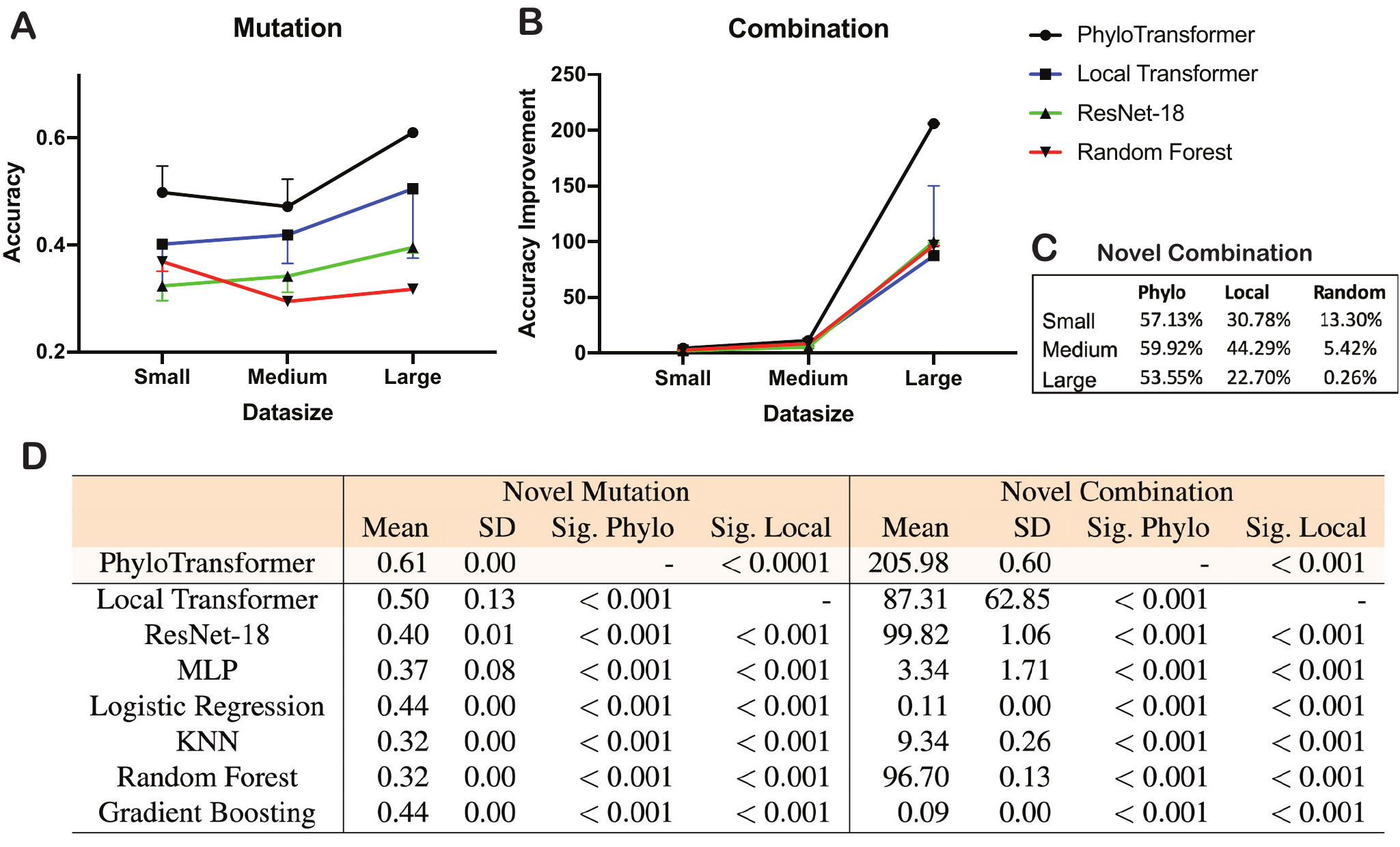}
\end{figure}

\noindent \bf Box 1 | Prediction Accuracy. A. \rm Prediction accuracy of \it novel mutations \rm from the \it small, medium, \rm and \it large \rm datasets based on PhyloTransformer and the best baseline methods. \bf B. \rm Prediction accuracy of \it novel combinations \rm trained with the \it small, medium, \rm and \it large \rm datasets based on PhyloTransformer and the best baseline methods. The accuracy improvement for each indicated model was calculated based on dividing the number of correct predictions by the expected number of correct random guesses. \bf C. \rm Prediction accuracy achieved by PhyloTransformer and Local Transformer averaged over 10 checkpoints after convergence. The expected prediction accuracy of random guesses refers to the average of 10 random guessing trials. \bf D. \rm Prediction accuracy of PhyloTransformer- and baseline method-generated predictions of \it novel mutations \rm and \it novel combinations. \rm Sig.~Phylo: $p$-value with respect to PhyloTransformer. Sig.~Local: $p$-value with respect to Local Transformer. 

\subsection*{Predicting Novel Combinations}
If a sequence in the testing set does not exist in the training set, we compared it to the reference sequence, then masked the mutated positions and generated predictions at these positions. If the model predicts \it all \rm the mutations correctly in this sequence, we say that it predicted a \it novel combination \rm correctly. The accuracy of predicting \it novel combinations \rm is the proportion of the number of sequences whose combinations are predicted correctly to all the sequences in the testing set.

\rm The difficulty of predicting \it novel combinations \rm changes as the size of the dataset changes, so we measure our prediction efficacy by accuracy improvement, which is defined as the following:
$$
\text{Accuracy improvement of a model} := {\text{Accuracy of the model} \over \text{Accuracy of random guessing}}.$$
For the \it small \rm dataset, there were 2.26 mutations on average with a standard deviation (SD) = 5.06; 
for the \it medium \rm dataset, there were 3.06 mutations on average with an SD = 2.56;
and for the \it large \rm dataset, there were 8.75 mutations on average with an SD = 2.87. 
For the \it small \rm dataset, random guessing resulted in an accuracy of 13.30\% with an SD = 1.12\%;
for the \it medium \rm dataset, random guessing resulted in an accuracy of 5.42\% with an SD = 0.12\%;
and for the \it large \rm dataset, random guessing resulted in an accuracy of 0.26\% with an SD = 0.012\%. The predicted results are summarized in Box 1B, where the accuracy improvement value was defined as follows: given the dataset (\it small, medium, \rm or \it large\rm), take the number of correct predictions generated by the indicated model and divide that value by the expected number of correct random guesses.

We performed a two-sample $z$-test of proportions to determine whether the accuracy of predicting \it novel combinations \rm by PhyloTransformer significantly less than baseline models on the \it large \rm dataset. The prediction accuracy of PhyloTransformer among the 10 checkpoints was higher than that generated by all of the baseline models with statistical significance. Local Transformer was no longer the best baseline model, while ResNet-18 and random forest outperformed Local Transformer for the task of predicting \it novel combinations. \rm
\begin{table}[!htb]
\centering
\begin{tabular}{|c|c|c|c|c|ccc|ccc|}
\hline
\hline
\rowcolor{orange!20}& \multicolumn{3}{c|}{In Amino Acid} &\multicolumn{7}{c|}{In Nucleotide} \\
\hline
\rowcolor{orange!5}Ranking & Location & Ref Seq & Predicted & Location & \multicolumn{3}{c}{Ref Seq} & \multicolumn{3}{|c|}{Predicted}\\
\hline
1 & 587 & I & T & 1759 & A & \textcolor{red}{T} & T & A & \textcolor{red}{C} & T\\
\rowcolor{orange!5}2 & 742 & I & T & 2224 & A & \textcolor{red}{T} & T & A & \textcolor{red}{C} & T\\
3 & 538 & C & R & 1611 & \textcolor{red}{T} & G & T & \textcolor{red}{C} & G & T\\
\rowcolor{orange!5}4 & 1080 & A & V & 3238 & G & \textcolor{red}{C} & C & G & \textcolor{red}{T} & C\\
5 & 720 & I & T & 2158 & A & \textcolor{red}{T} & T & A & \textcolor{red}{C} & T\\
\rowcolor{orange!5}6 & 851 & C & R & 2550 & \textcolor{red}{T} & G & T & \textcolor{red}{C} & G & T\\
7 & 423 & Y & H & 1266 & \textcolor{red}{T} & A & T & \textcolor{red}{C} & A & T\\
\rowcolor{orange!5}8 & 377 & F & S & 1129 & T & \textcolor{red}{T} & T & T & \textcolor{red}{C} & T\\
9 & 823 & F & L & 2466 & \textcolor{red}{T} & T & C & \textcolor{red}{C} & T & C\\
\rowcolor{orange!5}10 & 488 & C & R & 1461 & \textcolor{red}{T} & G & T & \textcolor{red}{C} & G & T\\
11 & 819 & E & G & 2455 & G & \textcolor{red}{A} & A & G & \textcolor{red}{G} & A\\
\rowcolor{orange!5}12 & 617 & C & F & 1849 & T & \textcolor{red}{G} & C & T & \textcolor{red}{T} & C\\
13 & 749 & C & R & 2244 & \textcolor{red}{T} & G & C & \textcolor{red}{C} & G & C\\
\rowcolor{orange!5}14 & 873 & Y & H & 2616 & \textcolor{red}{T} & A & C & \textcolor{red}{C} & A & C\\
15 & 1059 & G & V & 3175 & G & \textcolor{red}{G} & T & G & \textcolor{red}{T} & T\\
\rowcolor{orange!5}16 & 539 & V & A & 1615 & G & \textcolor{red}{T} & C & G & \textcolor{red}{C} & C\\
17 & 421 & Y & H & 1260 & \textcolor{red}{T} & A & T & \textcolor{red}{C} & A & T\\
\rowcolor{orange!5}18 & 877 & L & P & 2629 & C & \textcolor{red}{T} & G & C & \textcolor{red}{C} & G\\
19 & 418 & I & T & 1252 & A & \textcolor{red}{T} & T & A & \textcolor{red}{C} & T\\
\rowcolor{orange!5}20 & 1145 & L & S & 3433 & T & \textcolor{red}{T} & A & T & \textcolor{red}{C} & A\\
\hline
\hline
\end{tabular}
\caption{\textbf{Top 20 \textit{novel mutations} predicted by training PhyloTransformer with the \textit{large} dataset.} Ref Seq:  reference sequence hCoV-19/Wuhan/WIV04/2019 sequence (WIV04).}\label{tab:top_predictions}
\end{table}

\subsection*{Predictions in the Spike Protein RBM }
SARS-CoV-2 infects human cells by binding of the viral surface protein spike to its receptor on human cells, the ACE2 protein. Because of its role in viral entry, the RBD is a dominant determinant of zoonotic cross-species transmission. Although SARS-CoV-2 does not cluster within SARS and SARS-related coronaviruses, the RBD of SARS-CoV and SARS-CoV-2 share structural similarities, probably due to their shared zoonotic ancestry. This similarity implies convergent evolution for improved binding to ACE2 between the SARS-CoV and SARS-CoV-2 RBDs. Therefore, we focused our predictions on the spike protein RBD. The total length of the SARS-CoV-2 spike protein is 1,273 amino acids, and its structural features are listed below:
\begin{itemize}
    \item A signal peptide is located at the N-terminus (1–13 residues).
    \item The S1 subunit (14–685 residues) is responsible for receptor binding. The S1 subunit contains an N-terminal domain (14–305 residues), a C-terminal domain 0 (306-330 residues), an RBD (331-527 residues), a C-terminal domain 1 (528-590 residues), and a C-terminal domain 2 (591-685 residues).
   \item 	The S2 subunit (686–1273 residues) is responsible for receptor binding and membrane fusion. The S2 subunit contains cleavage sites (686-815 residues) at S1/S2 and S2', a fusion peptide (816–855 residues), a fusion peptide region (856-911 residues), a heptapeptide repeat sequence 1 (912–984 residues), a center helix (985-1034 residues), a connector domain (1035-1080 residues), a connector domain 1 (1081-1147 residues), a heptapeptide repeat sequence 2 (1163–1213 residues), a transmembrane domain (1213–1237 residues), and a cytoplasmic domain (1237–1273 residues) \cite{huang2020structural}. 
\end{itemize}

The spike protein RBM comprises amino acids 438 to 506. Yi et al.~\cite{yi2020key} compared the SARS-CoV-2 and SARS-CoV RBD affinity for hACE2 by creating single amino acid substitution mutations in the SARS-CoV and SARS-CoV-2 RBM sequences. The authors found that receptor binding was enhanced by introducing amino acid changes at P499, Q493, F486, A475, and L455, which are all localized to the RBM. PhyloTransformer trained with the \it large \rm dataset predicted only two mutations. The first mutation was predicted at amino acid 488, changing it from C to R, which is closely adjacent to F486. The second mutation was predicted at amino acid 497, changing it from F to S, once again right next to P499. The close proximity of the introduced mutations and predicted mutations indicated that PhyloTransformer is potentially capable of capturing meaningful genetic phenomena and can generate effective predictions. Our prediction results are reported in Table \ref{tab:pred}.
\begin{table}
\centering
\begin{tabular}{|c|c|c|c|ccc|ccc|}
\hline
\hline
\rowcolor{orange!20} \multicolumn{3}{|c|}{In Amino Acid} &\multicolumn{7}{c|}{In Nucleotide} \\
\hline
\rowcolor{orange!5}Location & Ref Seq & Predicted & Location & \multicolumn{3}{c|}{Ref Seq} & \multicolumn{3}{c|}{Predicted}\\
\hline
488 & C & R & 1461 & \textcolor{red}{T} & G & T & \textcolor{red}{C} & G & T\\
497&F & S&1489 & T & \textcolor{red}{T} & C & T & \textcolor{red}{C} & C\\
\hline
\hline
\end{tabular}
\caption{\textbf{Prediction of spike protein RBM mutations.} Ref Seq: reference sequence hCoV-19/Wuhan/WIV04/2019 sequence (WIV04).}\label{tab:pred}
\end{table}

\subsection*{Prediction of Glycosylation Site Modifications}

\begin{figure}[!htb]
     \centering
     \includegraphics[width=\textwidth]{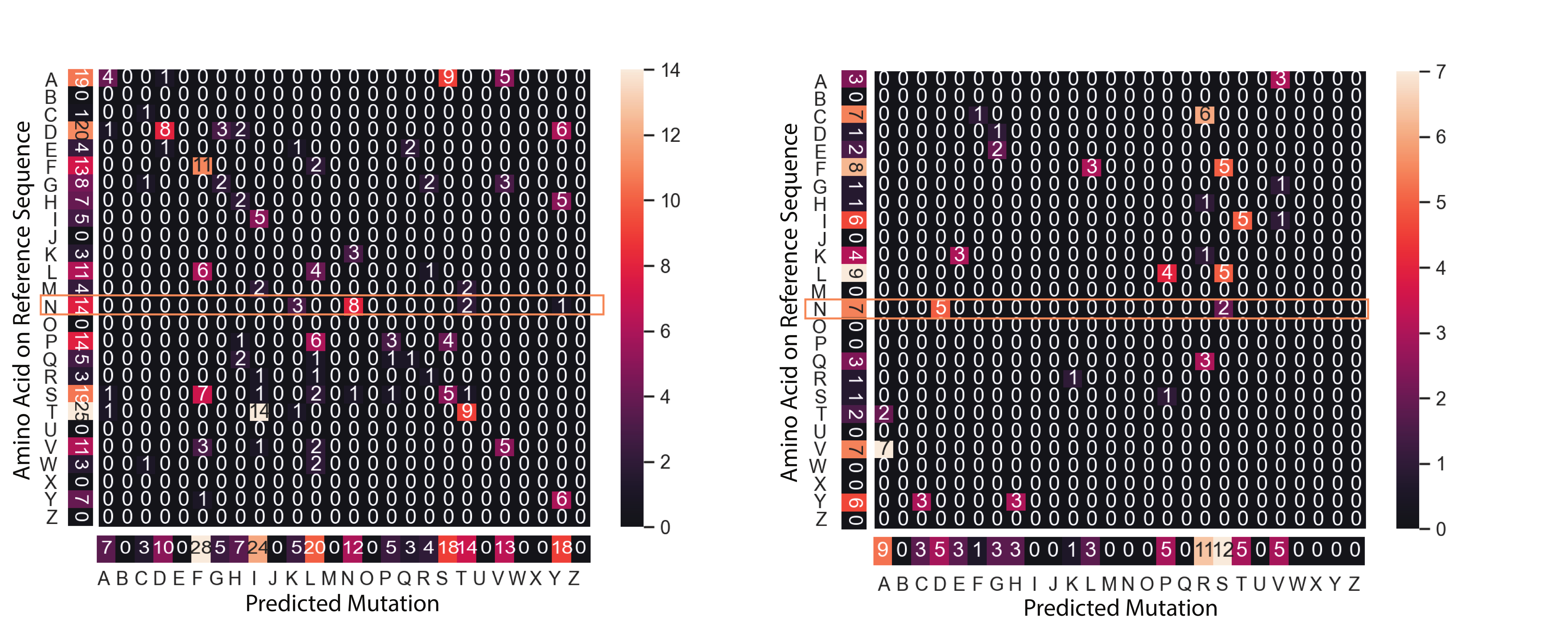}
     \caption{\bf Predicting amino acid changes. \textcolor{black}{\bf Left:} \rm existing mutations. \textcolor{black}{ \bf Right:} \rm predicted \it novel mutations \rm (total: 196 predictions). \textbf{Row:} amino acids of the SARS-CoV-2 reference sequence (total: 69 predictions). \textbf{Column:} predicted amino acids.}
     \label{fig:heatmap}
 \end{figure}

The SARS-CoV-2 spike protein is heavily glycosylated. Viral glycosylation plays a vital role in viral pathobiology, including antibody resistance, target recognition, viral entry, and host immune modulation \cite{doores2015hiv}. Glycosylation sites facilitate immune evasion by shielding epitopes from antibody neutralization; therefore, they are under selective pressure. 
Since glycosylation site modifications of the SARS-CoV-2 spike protein will likely impact the overall activities of SARS-CoV-2 replication and escape from immune surveillance \cite{hoffmann2021identification}, we examined glycosylation site model predictions. We reported our results on the N-glycosylation sites to help identify mutations associated with altered glycosylation that are favored during viral evolution. PhyloTransformer predicted three mutations of the following glycosylation sites: N122, N331, and N343. Table \ref{tab:gly} shows the predicted mutations in the spike protein changing N to a different amino acid. Figure \ref{fig:heatmap} summarizes the predicted mutations, including existing mutations (left) and \it novel mutations \rm (right), with predictions mutating away from amino acid N highlighted.

\begin{table}[!htb]
\centering
\begin{tabular}{|c|c|c|c|c|ccc|ccc|}
\hline
\hline
\multirow{2}{*}{Sites} &
\multicolumn{3}{c|}{In Amino Acid} &\multicolumn{7}{c|}{In Nucleotide} \\
\cline{2-11}
 & Location & Ref Seq & Predicted & Location & \multicolumn{3}{c|}{Ref Seq} & \multicolumn{3}{c|}{Predicted}\\
\hline
\rowcolor{orange!20} Glycosylation & 122 & N & D & 363 & \textcolor{red}{A} & A & C & \textcolor{red}{G} & A & C\\
\rowcolor{orange!20}Glycosylation & 331 & N & D & 990 & \textcolor{red}{A} & A & T & \textcolor{red}{G} & A & T\\
\rowcolor{orange!20}Glycosylation & 343 & N & S & 1027 &A&\textcolor{red}{A}&C & A&\textcolor{red}{G}&C\\
\hline
\rowcolor{orange!5}N-mutation  & 422 & N & D & 1263 & \textcolor{red}{A} & A & T & \textcolor{red}{G} & A & T\\
\rowcolor{orange!5}N-mutation  & 542 & N & D & 1623 & \textcolor{red}{A} & A & C &  \textcolor{red}{G} & A & C\\
\rowcolor{orange!5}N-mutation  & 542 &  N & S & 1624 & A & \textcolor{red}{A} & C & A & \textcolor{red}{G} & C\\
\rowcolor{orange!5}N-mutation  & 953 & N & D & 2856 & \textcolor{red}{A} & A & C & \textcolor{red}{G} & A & C\\
\hline
\hline
\end{tabular}
\caption{\textbf{Predictions of glycosylation sites and N-mutations.} \it First three rows: \rm predicted glycosylation site mutations. \it N-mutation sites: \rm other predictions with mutations of N. Ref Seq: reference sequence hCoV-19/Wuhan/WIV04/2019 sequence (WIV04).}\label{tab:gly}
\end{table}

\section*{Methodology} \label{sec:method}
\subsection*{Technical Background}
In this section, we will briefly review the history of sequence models that led to the development of Transformer and then introduce our PhyloTransformer model. The recurrent neural network (RNN) is the standard neural sequence model which extends the conventional feed-forward neural network with a recurrent hidden state dependent on the previous timestep. RNN  and its variants, such as the long short-term memory (LSTM) \cite{hochreiter1997long} and the gated recurrent unit (GRU) \cite{cho2014properties}, have been widely applied to important AI tasks, including language modeling \cite{mikolov2010recurrent}, speech recognition \cite{graves2013speech},  handwriting recognition \cite{graves2008novel}, and machine translation \cite{kalchbrenner2013recurrent}. However, RNNs are difficult to train in practice since the gradients tend to either vanish or explode as the sequence length increases \cite{bengio1994learning}. In addition, these models encode a source sequence into a fixed-length vector, which becomes a bottleneck when tackling particularly long sequences. 
Therefore, the attention mechanism was introduced \cite{bahdanau2014neural} to augment RNNs with an additional variable-length representation when encoding the input sequence. The attention mechanism allows the model to only focus on a subset of the input sequence for decoding.
The Transformer model comprises a purely attention-based network architecture without RNN backbones to directly capture intra-position dependencies via the self-attention mechanism \cite{vaswani2017attention}.
In self-attention, each sequence item has direct access to all the other positions, which yields a more powerful global representation of the sequence. This feature also inspires biological applications due to the long-range interactions of genetic sequences. 
However, the following challenges in modeling mutations on RNA sequences remain:
\begin{itemize}
    \item \bf Length adaptation: \rm most natural language processing (NLP) models deal with sequence lengths of a few hundred to a thousand, but the RNA sequence of SARS-CoV-2 is much longer: the genome of SARS-CoV-2 is 29,903 nucleotides in length \cite{kim2020architecture}, and the spike protein has 3,819 nucleotides.
    \item \bf Mutation sparsity: \rm due to the proofreading functions of coronaviruses \cite{smith2013coronaviruses}, mutations in the SARS-CoV-2 genome are rare. Our dataset shows consistency in this regard.
\end{itemize}

Regular Transformer scales quadratically with respect to the input sequence length, and the sparsity of mutations might lead to the generative Transformer model overfitting the identical parts while ignoring the mutations. Therefore, to adapt to biological problems and address issues regarding genetic mutations, a new model that tackles the length and sparsity issues commonly encountered in existing deep neural network architectures is required. To address these two challenges, we propose PhyloTransformer, which is a linear time complexity discriminative model based on the Transformer architecture. The time and space linearity are achieved by adopting FAVOR+ from Performer \cite{choromanski2021rethinking}, which performs an unbiased fast attention approximation with low variance. The mutation sparsity issue is addressed by adopting the MLM training objective from BERT \cite{devlin-etal-2019-bert}, which is a discriminative variant of Transformer for supervised NLP tasks. 
A detailed description of PhyloTransformer architecture is presented in the next section.
\subsection*{Model Development}
We adopted a discriminative approach to model the mutation probability at a particular position in the RNA sequence. Let $p(x_i =A | X)$ denote the probability of the $i^{\textrm{th}}$ nucleotide changing to $A$ given the reference sequence $X$. We will demonstrate how to predict $p(x_i | X)$ by PhyloTransformer and other baseline models in this section.

\subsubsection*{The PhyloTransformer Model}
The PhyloTransformer model adopts a Transformer-based network, which utilizes the full spike sequence of 3,819 nucleotides as input and generates output mutation probabilities at particular positions. We followed the MLM pre-training objective from BERT \cite{devlin-etal-2019-bert}. 
Note that the attention mechanism in Transformer \cite{vaswani2017attention} calculates attention matrices with a shape of $L \times L$ (where $L$ is the length of the sequence) to capture the relationship between nucleotides. In order to reduce the computation complexity of the attention matrix, we adopted the FAVOR+ technique from Performer \cite{choromanski2021rethinking}, which 
performs approximate attention computation in linear time. In the following content, we first present the network architecture of PhyloTransformer. Next, we introduce FAVOR+ for fast low-rank approximation of the regular full-rank attention computation in linear time. Finally, the overall training process will be discussed in detail.
\paragraph{Bidirectional Transformer Encoder: }
Let $X = (x_1, x_2, ..., x_L)$ denote the reference sequence, where $x_i$ is the nucleotide at position $i$ in the RNA sequence. 
We first applied trainable projections to map each $x_i$ with its position information to three embedding vectors, $q_i$, $k_i$ and $v_i$, for attention computation. 
Suppose the dimension of each embedding is $d$. The output of the attention layer is computed by the following equation:
\begin{equation}
\label{raw_attention}
    \mathrm{Attention}(Q, K, V) = A\cdot V = \mathrm{softmax}\left(\frac{QK^{T}}{\sqrt{d}}\right)V
\end{equation}
where $A\in\mathbb{R}^{L\times L}$ is the attention matrix. $Q = [q_1; q_2; ...; q_L], K = [k_1; k_2; ...; k_L]$, and $V = [v_1; v_2; ..., v_L]$ are embedding matrices in $\mathbb{R}^{L\times d}$, where $q_i, k_i$, and $v_i$ are row vectors representing three embeddings. After the attention layer is computed, we further applied a feed-forward layer with a residual connection. An attention layer and a feed-forward layer compose a single Transformer module. We stacked the $N$ Transformer modules as the overall network architecture of our PhyloTransformer model. 
%
%
\paragraph{FAVOR+: }
In the original attention mechanism, the time complexity of computing the attention layer by Equation (\ref{raw_attention}) is $O(L^2{d})$, which becomes computationally intractable when $L$ is large. The Performer \cite{choromanski2021rethinking} model proposed kernelizable attention by deriving a mapping $\phi$ to decouple the attention matrix $A$ into $Q'$ and $K'$, where $q_i' = \phi(q_i), k_i'=\phi(k_i)$ and $Q', K' \in \mathbb{R}^{L\times r}, r \ll L$. In this case, the attention layer can be computed by the following equation:
\begin{equation}
    \label{fast_attention}
    \mathrm{Attention}(Q, K, V) = D^{-1}(Q'((K')^{T}V)), \quad D = \mathrm{diag}(Q'((K')^{T}\boldsymbol{1}_L))
\end{equation}
where $\boldsymbol{1}_L$ is an all-ones vector of length $L$. Since $Q', K' \in \mathbb{R}^{L\times r}, V \in \mathbb{R}^{L\times d}$, the computation complexity decreases to $O(rLd)$ with respect to a small constant $r$, making it computationally feasible to handle particularly long sequences such as RNA data. 

\paragraph{Training process:} 
We denoted the reference sequence by $X = (x_1, x_2, ..., x_L)$ and the mutated sequence by $Y = (y_1, y_2, ..., y_L)$. During the training process, we masked some positions in $X$, and used the model to predict nucleotides in $Y$ at those masked positions. 
Fig.~\ref{fig:mlm_model} shows the workflow of our model. Specifically, we first identified the set of mutated positions $\mathcal{P}_m = (P_{1}, \ldots, P_{k})$, where $P_1, \ldots, P_k$ are indices. 
In addition, we also randomly chose some unchanged positions 
$\mathcal{P}_u = (P'_{1}, \ldots, P'_{r})$ such that $\frac{\left|\mathcal{P}_m \cup \mathcal{P}_u \right|}{L} = 0.015$. Next, we applied a masking function $f_m(x_i)$ to each nucleotide $x_i$ at the masking positions. Namely, $\forall P_i \in \mathcal{P}_m \cup \mathcal{P}_u$, we have:
\begin{figure}[ht]
    \centering
    \includegraphics[width=0.6\linewidth]{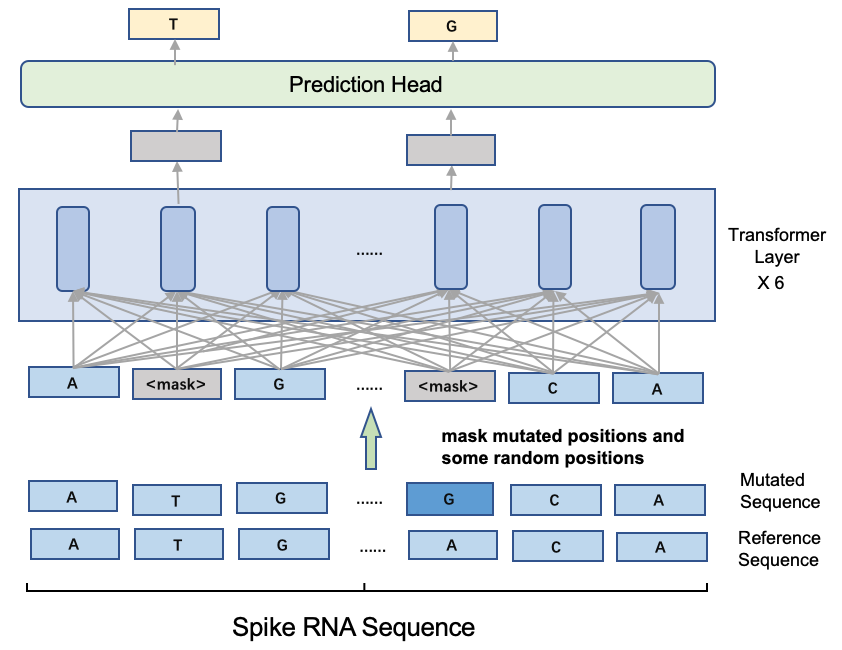}
    \caption{\bf The training scheme of PhyloTransformer. \rm We compared the reference sequence to a mutated sequence and focused on the mutations. We masked the mutated positions along with some random positions and processed the masked sequence with stacked Transformer modules. The final hidden state of the stacked Transformer layer for each masked position was used as the aggregate representation for the mutation prediction task.} \label{fig:mlm_model}
\end{figure}
\begin{equation}
    \label{mask_function}
    f_m(x_i) = \left\{
    \begin{aligned}
        & <mask> & 80\% \quad \text{of cases} \\
        & x_i & 10\% \quad \text{of cases} \\
        & \mathrm{Random}(\{A, T, C, G\}) & 10\% \quad \text{of cases} \\
    \end{aligned}
\right.,
\end{equation}
where $<mask>$ is a special masking token. The masking function $f_m$ acts on $1.5\%$ of the entire nucleotides and further randomly maps each nucleotide from this masking subset to a special token  $ <mask>$ (80\% chance), a random substitution (10\%) or itself (10\%). Denoting the masked sequence as $\widetilde{X}$, we encode $\widetilde{X}$ with stacked Transformer modules and represent each nucleotide as a hidden vector $h_i$. Next, the probability of $y_i$ at each masking position is computed as follows:
\begin{equation}
    \label{prob}
    P(y_i | \widetilde{X}) = \mathrm{softmax}(W_o h_i)\quad\forall i\in \mathcal{P}_m \cup \mathcal{P}_u,
\end{equation}
where $W_o$ are trainable parameters. The probability of all the masked nucleotides is the following equation:
\begin{equation}
    \label{prob_all}
    P(Y | \widetilde{X}) = \prod_{i \in \mathcal{P}_m \cup \mathcal{P}_u} P(y_i | \widetilde{X})
\end{equation}
The model is optimized to minimize the negative log probability over all the mutated sequences from the training set $\mathcal{Y}$ with respect to different masking positions, as determined by the equation:
\begin{equation}
    \label{loss}
    L(\theta) = - \sum_{Y \in \mathcal{Y}}\mathbb{E}_{f_m}\left[\log P(Y | \widetilde{X})\right].
\end{equation}
\begin{equation}
    \label{loss}
    L(\theta) = - \mathbb{E}_{Y \in \mathcal{Y}} \left\{ \mathbb{E}_{i \in \mathcal{P}_m \cup \mathcal{P}_u}\left[\log P(y_i | \widetilde{X})\right] \right\}.
\end{equation}
Since most of the masked positions are mutated positions,  our model is trained to concentrate on mutation predictions. Meanwhile, the randomly chosen positions (i.e., $\mathcal{P}_u$) also improved the robustness of our model.

\subsubsection*{Local models}
\begin{figure}
    \centering
    \includegraphics[width=0.85\linewidth]{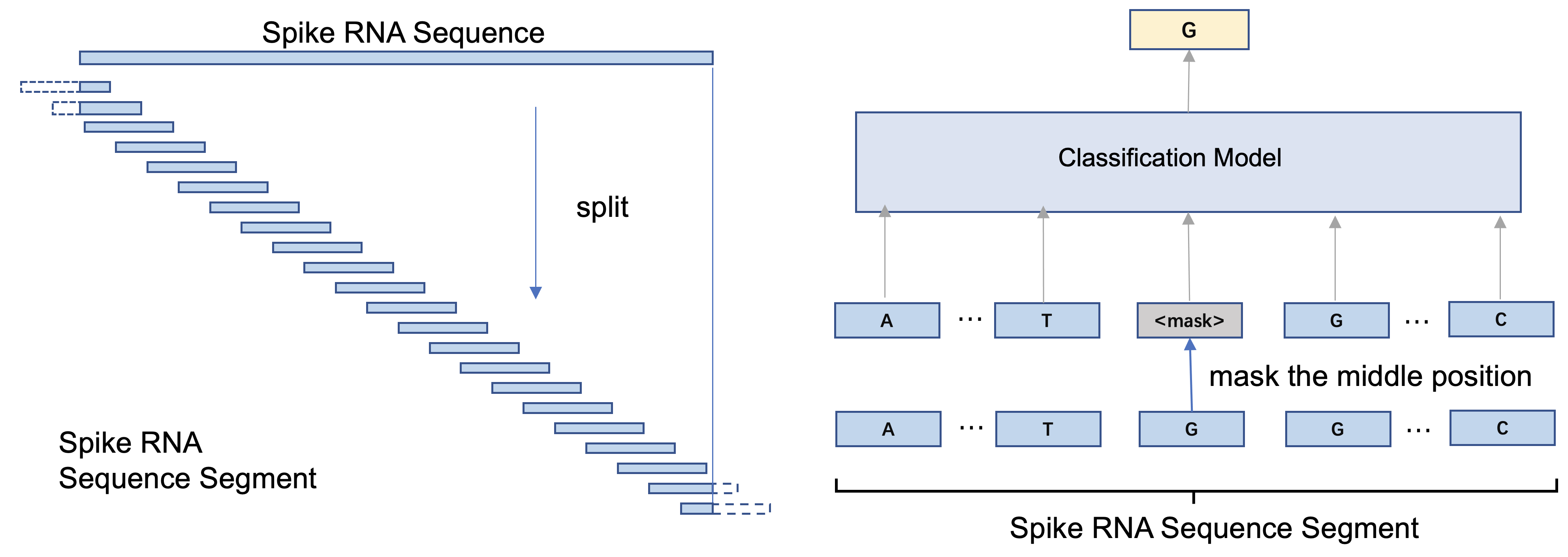}
    \caption{\bf The training scheme of local models. \rm The spike RNA sequence was split into 3,819 segments with padding, and the middle nucleotide was masked in each segment. A classification model was adopted to predict the masked nucleotides.}
    \label{fig:split_transformer}
\end{figure}

In addition to PhyloTransformer, which considers the full sequence, we also examined baseline methods, which predict $p(x_i | X)$ based on local segments from the spike RNA sequence. 
There is a total of 3,819 nucleotides in the spike sequence. 
We can obtain a local segment of 15 nucleotides centered around each nucleotide with sequence padding. Thus, we can obtain 3,819 segments of 15 nucleotides from the full spike RNA sequence. The center position of each segment is masked. We adopted various classification methods (including neural models and non-neural methods) to predict the center nucleotide based on other nearby nucleotides. During the training phase, we split all training spike RNA sequences into segments and generated a local dataset with repeated segments filtered out. The training process is shown in Figure~\ref{fig:split_transformer}, where any classification method could be used, such as the standard Transformer, ResNet-18, MLP, logistic regression, KNN, random forest, and gradient boosting.

\subsubsection*{Training details}
For the PhyloTransformer model, we stacked six Transformer modules with eight attention heads and a hidden size of $1,024$. We optimized the model following the loss function in Eq.~\ref{loss} with Adam ($\beta_1 = 0.9, \beta_2 = 0.999$). We chose a learning rate of $3e-5$ for all three datasets. The batch size was $16$ for the \it small \rm and \it medium \rm datasets, and the batch size was $32$ for the \it large \rm dataset. For the \it large \rm dataset, we trained the PhyloTransformer model \rm with 8 Nvidia V100 GPUs for 10 epochs, which took 13 hours per epoch.
For Local Transformer, we stacked $12$ Transformer modules with eight attention heads and a hidden size of $768$ for better representation capability. We employed a standard classification loss and optimized the model via Adam ($\beta_1=0.9, \beta_2 = 0.98$). We used the learning rate of $1e-4$ with a batch size of $128$ for all three datasets on a single Nvidia 3080 GPU with 100 training epochs. For the \it large \rm dataset,  each epoch was completed in approximately 20 minutes. 
For the ResNet model, a popular variant of the convolutions neural network, we employed the ResNet-18 architecture as our backbone, and Adam ($\beta_1 =0.9, \beta_2 = 0.999$) was utilized as the optimizer with a learning rate of $5e-5$, and a batch size of $128$. We trained the ResNet model for 100 epochs on a single Nvidia 3080 GPU.
 In the \it large \rm dataset, a single epoch 
 was completed in approximately five minutes. For other methods, we used scikit-learn (0.23.2) \cite{pedregosa2011scikit} with its default settings.

\section*{Discussion}

\subsection*{Significance}
The overall goal of our research is to train a state-of-the-art sequence model using existing viral genetic sequence data to identify SARS-CoV-2  variants that may have evolutionary advantages and become the emerging VOCs. In this paper, we developed the PhyloTransformer model, a novel deep neural network with a multi-headed self-attention mechanism.
PhyloTransformer was subjected to an 
advanced training methodology to predict potential mutations that may lead to enhanced virus transmissibility or resistance to antisera. 
Our computational platform may be helpful in guiding the design of therapeutics and vaccines for effective targeting of emerging SARS-CoV-2 VOCs, as well as novel mutants of other viruses that may cause pandemics.

\subsection*{Technical Insight}
Compared to baseline models that only utilize local information, PhyloTransformer aggregates the information from the full sequence. The Transformer backbone of the PhyloTransformer model \rm utilizes the self-attention mechanism to capture the relationships between nucleotides over the whole sequence. By comparing the results obtained with PhyloTransformer and Local Transformer, we observed that modeling the intra-sequence relationship for long sequences is beneficial. 
However, direct modeling of long sequences is costly.
The FAVOR+ technique from Performer \cite{choromanski2021rethinking} reduces the time and space complexity of training from quadratic to linear. 
This reduction in computational complexity is critical for processing RNA sequences. Meanwhile, the MLM training objective adopted by PhyloTransformer enables a discriminative formulation that resolves various technical issues and allows for practical biological applications. 
In general, we believe that the capability of modeling long sequences with computational efficiency and representing biological characterizations more accurately may be the key to more reliable predictions.
\subsection*{Future Work}

Our future studies will focus on the application of bioinformatics, phylogeny, and spatial information to the study of the viral envelope for enhanced prediction performance. We also aim for bench validation of these mutations with pseudoviral models of SARS-CoV-2 to access (1) the infectivity of predicted spike mutants of emerging concern; (2) resistance to antisera; and (3) enhancement of mutant spike cell-cell fusion. We hope that the predicted mutations may serve as early identification of potential VOCs or variants of high consequence that will not only be a valuable resource for researchers studying SARS-CoV-2 but may also allow for the creation of more robust vaccines and therapies for the treatment of COVID-19 in the future.

\vskip 5mm

\noindent\textbf{Acknowledgements}\\
The computations in this paper were run on the FASRC Cannon cluster supported by the FAS Division of Science Research Computing Group at Harvard University.\\

\noindent\textbf{Author contributions}\\
Y.W., S.-T.Y., and Y.W.~conceived the project, Y.W.~and Y.W.~designed the model architecture and training paradigm, Y.W.~and S.X.~developed the prediction models, and Y.W.~performed the mathematical analysis. Y.W.~and S.X.~performed the numerical experiments. Y.W.~wrote the paper. All authors contributed to all aspects of the project.\\

\noindent\textbf{Competing interests}\\
The authors declare no competing interests.

\bibliography{sample}

\end{document}